\documentstyle[multicol,aps]{revtex}
\def\baselinestretch{1.1}

\renewcommand{\narrowtext}{\begin{multicols}{2} \global\columnwidth20.5pc}
\renewcommand{\widetext}{\end{multicols} \global\columnwidth42.5pc}

\begin{document}

\title{\Large\bf  Wilson lines on noncommutative tori }
\author{ \large\bf Anton Yu.~Alekseev$^{\dag}$,\ \
 Andrei G.~Bytsko$^*$ \\[2mm] }

\address{\noindent $^{\dag}$  
Institute for Theoretical Physics, Uppsala University,
 Box 803, S-75108, Uppsala, Sweden\\
  alekseev@teorfys.uu.se  \\[1mm] }
 
\address{ $^*$ Steklov Mathematics Institute,
Fontanka 27, St. Petersburg 191011, Russia\\
bytsko@pdmi.ras.ru  \\[2mm] {\rm hep-th/0002101} }
\date{February, 2000} 
\maketitle

\begin{abstract}
\noindent

We introduce the notion of a monodromy
for gauge fields with vanishing curvature on the 
noncommutative torus. Similar to the ordinary
gauge theory, traces of the monodromies
define noncommutative Wilson lines. Our main
result is  that  these Wilson lines are 
invariant under the Seiberg-Witten map changing the 
deformation parameter of the noncommutative torus.

\vspace{1mm}
\noindent
PACS-98: 11.15.-q, 11.10.Kk, 02.90.+p
\end{abstract}

\narrowtext

\section{Introduction}
Gauge theories on noncommutative spaces \cite{NG} arise
as low energy effective theories on $D$-brane world
volumes in the presence of the background $B$-field in the theory 
of open strings \cite{DH,VS,SW}. The simplest
noncommutative space where a gauge theory can be
constructed is the space ${\bf R}^{d}_\theta$ with
coordinates $x^i$ obeying the Heisenberg commutation
relations,
\begin{equation} \label{xx}
  [x^i, x^j] = i \, \theta^{ij} \,,
\end{equation}
where $\theta^{ij}=-\theta^{ji}$ is a constant
anti-symmetric matrix. In string theory, $\theta$ 
is given by the following formula \cite{VS,SW}
$$
 \theta=-(2\pi \alpha')^2(g+2\pi \alpha'B)^{-1} B 
         (g-2\pi \alpha'B)^{-1} \,,
$$
where $\alpha'$ is the inverse to the string tension,
$g$ is the metric and $B$ is the $B$-field
on the brane world-volume.
Functions on the space ${\bf R}^{d}_\theta$ can be
identified with ordinary functions on ${\bf R}^{d}$
with the noncommutative product given by the Moyal formula,
\begin{equation} \label{Mo}
 (u  * v)(x) = \Bigl( e^{ \frac i2 \theta^{ij}  
 \partial_i^x \partial_j^y }\, u(x) v(y) \Bigr)_{y=x} \,.
\end{equation}

Along with the space ${\bf R}^{d}_\theta$ it is natural to
consider its compactification $T^{d}_\theta$. We can choose
the compactification radii $R_i$ such that the
functions on $T^{d}_\theta$ are invariant with respect to
the shifts,
$$
f(x^1 + 2\pi n_1 R^1, \dots, x^d + 2\pi n_d R^d)=
f(x^1, \dots, x^d) \,,
$$
where $n_i$ are arbitrary integers. Since $\theta^{ij}$ is
a constant tensor, the Moyal product is invariant with respect
to such shifts. Therefore, the $*$-product restricts to periodic 
functions. 

An  important ingredient in the ordinary gauge theory
is the notion of a {\em Wilson line},
\begin{equation}  \label{Wilson}
W_\gamma(A) = {\rm Tr} \, P \exp( i\int_\gamma A),
\end{equation}
where $A$ is a gauge field and $\gamma$ is a closed contour.
The expression $W_\gamma(A)$ is gauge invariant and, hence,
defines an observable in the gauge theory. One reason why
$W_\gamma(A)$ is an important object is given by the 't Hooft's 
criterion of confinement \cite{tH}. 
Another reason is as follows. Suppose that the curvature of $A$,
$$
F_{ij} = \partial_i A_j - \partial_j A_i + i [A_j, A_i] \,,
$$
vanishes. Locally, the condition $F=0$ implies that $A$ is a pure 
gauge. Globally, flat connections (gauge fields with vanishing 
curvature) may have moduli. One can distinguish a nontrivial gauge 
field with $F=0$ from a pure gauge by looking at the Wilson lines 
along noncontractible contours. The standard application of this 
principle is the Aharonov-Bohm effect where the phase of the Wilson 
line for electromagnetic vector potential influences the interference 
pattern.

It is the goal of this paper to study the notion of a Wilson line 
in a gauge theory on the noncommutative torus $T^{d}_\theta$.
In contrast to the previous suggestions \cite{Wil,Ok}, we only consider
the case of flat connections. In our view, there is a conceptual
difficulty in the case of $F_{ij} \neq 0$: the definition of a Wilson 
line contains a contour $\gamma$. It might be a nontrivial task to 
find noncommutative subspaces of a noncommutative space. Technically 
speaking, the requirement is that the algebra of functions vanishing on 
the subspace be an ideal in the noncommutative algebra of functions.
For instance, if $\theta$ is nondegenerate and generic, the 
noncommutative torus $T^{d}_\theta$ has no subspaces different {}from 
$T^{d}_\theta$ itself. In particular, there is not a single 
{}`noncommutative contour' on $T^{d}_\theta$. In the case of $F_{ij}=0$, 
the Wilson line only depends on the homotopy class of the contour 
$\gamma$. We propose a natural gauge invariant object which generalizes 
$W_\gamma(A)$ in the case of flat noncommutative gauge fields.

It has been shown in \cite{SW} that one can construct a 
transformation (the Seiberg-Witten map) which identifies 
the noncommutative gauge fields on ${\bf R}^{d}_\theta$ 
(and on $T^{d}_\theta$) for different values of $\theta$. 
The Seiberg-Witten map on general noncommutative spaces
was recently studied in \cite{map}. 
We show that our noncommutative Wilson line is invariant with 
respect to the Seiberg-Witten map. In other words, we prove that 
the Seiberg-Witten map for flat connections is isomonodromic.

It would be very interesting to see whether the noncommutative
Wilson lines have some natural interpretation in the $D$-brane
Physics. The following elementary example can be used as a starting 
point. The energy spectrum of a non-relativistic particle of mass $m$ 
and charge $q$ confined to a circle of length $L$ in the constant 
vector potential $A$ has the form,
$$
E_n = \frac{1}{2m} \ \frac{4\pi^2 \hbar^2}{L^2} 
 \bigl( n + i  \ln(M) \bigr)^2 \,,
$$ 
where
$$
M= \exp \Bigl( i \ \frac{q A L}{2\pi \hbar} \Bigr)
$$
is the monodromy of the Wilson line winding around the circle.
In the case of noncommutative torus, the charged particle is replaced
by the end-point of an open string. It is plausible that the 
noncommutative Wilson line shifts the spectrum of an open string in 
a similar fashion.

\section{Noncommutative Wilson lines}
In ordinary gauge theory, there are two ways to define a Wilson 
line for a gauge field $A$ with vanishing curvature $F$.
For the purpose of this paper we restrict our attention
to the Wilson lines on the torus $T^d$.
Since a gauge field with vanishing curvature is locally a pure gauge,
one can solve the equation, 
\begin{equation}\label{dif}
 \partial_i g(x) = i A_i(x) \, g(x) \,, 
\end{equation}
where $g$ is not necessarily periodic. Different solutions of
(\ref{dif}) can be obtained from each other by multiplication
by a constant from the right, $g(x) \rightarrow g(x) h$.
Under  gauge transformations solutions of equation (\ref{dif})
get multiplied from the left, $g(x) \rightarrow h(x) g(x)$.

One can now introduce left and right monodromy matrices,
$M_i = g^{-1} g_i$ and $\tilde{M}_i = g_i g^{-1}$, where 
$g\equiv g(x^1, \dots, x^d)$ and $g_i \equiv g(x^1, \dots, x^i+2\pi R^i, 
\dots, x^d)$ are solutions of (\ref{dif}).
It is easy to see that the monodromies $M_i$ are gauge-invariant
and do not depend on the point $x$. If one replaces the solution
$g(x)$ by some other solution of equation (\ref{dif}), $g(x)h$, the 
monodromy $M_i$ gets conjugated, $M_i \rightarrow h^{-1} M_i h$. 
Therefore, the trace of $M_i$ is a gauge invariant object which
is independent of the choice of a solution of (\ref{dif}) and
can be used as the definition of the Wilson line, $W_i = {\rm Tr}\, M_i$.

The monodromies $\tilde{M}_i$ are the usual ordered exponentials,
$$
\tilde{M}_i = P \, \exp( i \int_{x^i}^{x^i+2\pi R^i} A ).
$$
They are independent of the choice of the solution $g(x)$ but
have explicit dependence on the point $x$ and transform covariantly
under the gauge transformations, 
$\tilde{M}_i \rightarrow h(x) \tilde{M}_i h(x)^{-1}$.
Again, taking a trace yields a gauge invariant object independent 
of all choices which coincides with $W_i$.

In the noncommutative gauge theory, only the prescription which
uses monodromies $M_i$ survives. It is still possible to construct
objects similar to $\tilde{M}_i$ (see {\em e.g.} \cite{Ok}) which 
transform covariantly under the gauge transformations, 
$\tilde{M}_i \rightarrow h(x) * \tilde{M}_i * h(x)^{-1}$. However, 
the matrix trace fails to be cyclic under the Moyal product, that 
is ${\rm Tr} (u*v) \neq {\rm Tr} (v*u)$. This implies that 
${\rm Tr} \, \tilde{M}_i$ {\em is not} gauge-invariant.

In what follows we present the definition of a Wilson line in
noncommutative gauge theory which uses the monodromy $M_i$.
In the noncommutative case, the curvature of the gauge field
is defined using the $*$-product (see {\em e. g.} \cite{SW}),
\begin{equation} \label{F}
 F_{ij} = \partial_i A_j - \partial_j A_i
 - i A_i * A_j + i A_j * A_i \, .
\end{equation}
The infinitesimal gauge transformations are of the form,
\begin{equation} \label{gA} 
 \delta_{\lambda} A_i = \partial_i \lambda + 
  i \lambda * A_i - i A_i * \lambda \,.  
\end{equation}
Let us mention that not every compact Lie group can be used as a gauge 
group on $T^d_\theta$. For instance, $G=U(N)$ works for any $\theta$ 
whereas $G=SU(N)$ in general does not give rise to a gauge group on the 
noncommutative torus. More precisely, the $*$-product $g*h$ of two 
unitary matrices is always unitary, but in general 
${\rm det} (g*h) \neq {\rm det}(g) * {\rm det}(h)$.

Equations (\ref{F}) and (\ref{gA}) are obtained from the standard 
formulas in ordinary (commutative) gauge theory by introducing 
$*$-products instead of the ordinary products. It is slightly more 
complicated to find a counterpart of the general formula for gauge 
transformations,
\begin{equation} \label{hA}
 A^h_i = - i \, \partial_i h * h_*^{-1} + 
  h * A_i * h_*^{-1} \,.  
\end{equation}
Here $h(x)$ is an element of the gauge group, and
$h_*^{-1}$ is the inverse of $h$ {\em with respect to}
the $*$-product, $h * h_*^{-1} = h_*^{-1} * h = 1$.
Note that for $\theta \neq 0$ the element $h_*^{-1}$
differs from the ordinary inverse $h^{-1}$.  More
explicitly,
$$
 h_*^{-1} = h^{-1} + \frac i2 \theta^{kl} h^{-1} 
 (\partial_k h) h^{-1} (\partial_l h) h^{-1} + {\cal O}(\theta^2) \,.
$$

We now turn to the case of a noncommutative gauge
field $A$ with vanishing curvature $F=0$. Then, one may 
expect that at least locally $A$ is a pure gauge. That is,
there is an element of the gauge group $g$ such that,
\begin{equation} \label{Ag}
  \partial_i g = i A_i * g \,.
\end{equation}
At this point it is convenient to view gauge fields 
on $T^d_\theta$ as periodic gauge fields on ${\bf R}^d_\theta$. 
Then, we expect that equation (\ref{Ag}) admits
global solutions. For now we assume that such solutions
exist and return to this issue in the next section.

Our next task is to show that the ratio $m=g_*^{-1} * g'$ of two
solutions $g$ and $g'$ of equation (\ref{Ag}) is a constant. Indeed,
\begin{eqnarray} \label{const}
 && \partial_i m  = -g^{-1}_* * \partial_i g
 *g^{-1}_* * g'+ g^{-1}_* * \partial_i g'  \nonumber \\ 
 && = -i\, g^{-1}_* * ( A_i - A_i ) *g' = 0 \,. 
\end{eqnarray}
Let us mention that in the derivation of this equation we have used 
that the $*$-product on ${\bf R}^d_\theta$ satisfies the Leibniz rule,
$\partial_i(f*g)=\partial_i f*g + f *\partial_i g$. In general, this 
property does not hold on noncommutative spaces.

If $g(x^1, \dots, x^d)$ is a solution of (\ref{Ag}) then
so is $g_i(x^1, \dots, x^d)=g(x^1, \dots, x^i+2\pi R^i, \dots, x^d)$.
By equation (\ref{const}), the ratio $M_i = g_*^{-1} * g_i$ is
independent of the point on $T^d_\theta$. We call
the group elements $M_i$ {\em noncommutative monodromies}
of the equation (\ref{Ag}). As usual, the value of the monodromy
depends on the choice of the solution $g(x)$. 
For some other solution, $g^\prime(x)=g(x)*h=g(x)h$ one gets,
$M^\prime_i=h^{-1}g_*^{-1} * g_ih=h^{-1}M_ih$. Here we used that 
the $*$-product with a constant coincides with the ordinary
product. Finally, we can define the Wilson lines which are
independent of the choice of the solution $g(x)$,
\begin{equation} \label{qW}
W_i(A)= {\rm Tr} \, M_i= {\rm Tr} \, (g_*^{-1} * g_i) \, .
\end{equation}
Let us stress, that, since $M_i$ are coordinate independent,
we use the ordinary matrix trace here.
Similar to their commutative counterparts (\ref{Wilson}), the
Wilson lines $W_i(A)$ are gauge invariant,
$$
W_i(A^h)={\rm Tr} \, (g_*^{-1}*h_*^{-1} * h * g_i) = W_i(A) \, ,
$$
where $g(x) \rightarrow h(x) * g(x)$ is the composition of 
two gauge transformations.

We put forward the definition (\ref{qW}) of noncommutative Wilson 
lines $W_i(A)$.  In the next Section we apply the Seiberg-Witten map 
to the noncommutative gauge theory at hand and show that our Wilson 
lines $W_i(A)$ are invariant with respect to this transformation.

Let us remark that although the considerations of this Section are 
restricted to the case of gauge fields with vanishing curvature, 
they may be viewed as the general definition of a noncommutative 
Wilson line. Indeed, in the commutative case the Wilson lines are 
assigned to closed contours. When restricted to a 1-dimensional contour, 
any gauge field has vanishing curvature and one can use the 
monodromy $M$ instead of the ordered exponential $\tilde{M}$ to 
construct the Wilson line. In the noncommutative context it is more 
difficult to construct submanifolds. 
In particular, as we pointed out in Introduction, the noncommutative
tori with generic $\theta$ have no nontrivial submanifolds. However, 
if one can find a subtorus $\Gamma$ of the noncommutative torus 
$T^d_\theta$ such that the restriction of the gauge field $A$ to 
$\Gamma$ has vanishing curvature, one can define the Wilson lines
of $A$ along $\Gamma$. If $\Gamma$ is 1-dimensional, it defines
one Wilson line observable, similar to  commutative gauge
theory.

\section{The Seiberg-Witten map}

It was established in \cite{SW} that the gauge theories on noncommutative
tori $T^d_\theta$ with different values of the deformation parameter
$\theta$ are equivalent to each other. More explicitly, let $\theta$ 
and $\theta+\delta \theta$ be two infinitesimally close values of the 
deformation parameter. Then, there exists the {\em Seiberg-Witten map} 
$A \rightarrow \hat{A}(A)$ and 
$\lambda \rightarrow \hat{\lambda}(\lambda,A)$ 
of the gauge fields and infinitesimal gauge parameters
on ${\bf R}^d_\theta$ to those on ${\bf R}^d_{\theta+\delta \theta}$
such that
\begin{equation} \label{map}
\hat{A}(A+\delta_\lambda A)=\hat{A}(A) + 
\delta_{\hat{\lambda}(\lambda, A)} \hat{A}(A) \, .
\end{equation}
This map is given by explicit formulas (see equation (3.8) in \cite{SW}).
In this paper we only need the form of the Seiberg-Witten
transformation for $A$ with $F=0$,
\begin{equation} \label{dA}
\delta_\theta A_i =  -\frac 14 \delta\theta^{kl} 
 ( A_k * \partial_l A_i + \partial_l A_i * A_k) \,.
\end{equation}
We show in the Appendix that under the Seiberg-Witten
map solutions of the equation (\ref{Ag}) transform
according to formula,
\begin{equation} \label{dg}
 \delta_\theta g = \frac 1{4i} \delta\theta^{kl} A_k * A_l * g \,.
\end{equation}
It is convenient to introduce a special notation for the
`gauge parameter',
\begin{equation}  \label{lambdatheta}
\lambda(\delta \theta, A) =  - \frac 1{4} \delta\theta^{kl} A_k * A_l \, .
\end{equation}
Note, however, that $\delta_\theta A_i$ {\em is not} a gauge 
transformation. More exactly,
$$
\delta_{\lambda(\delta \theta, A)} A_i =  -\frac 14 \delta\theta^{kl} 
 ( A_k * \partial_l A_i - \partial_l A_i * A_k) \,.
$$
The reason is that in contrast to partial derivatives $\partial_i$, 
the transformation $\delta_\theta$ violates the Leibniz rule for the 
Moyal product. Indeed, the exponential form of (\ref{Mo}) implies,
\begin{equation} \label{var}
 \delta_\theta (u*v) = (\delta_\theta u) *v +
 u* (\delta_\theta v) + \frac i2 \delta\theta^{ij} 
  \partial_i u * \partial_j v \,.
\end{equation}
In particular, putting together equations (\ref{dA})-(\ref{var}) 
we obtain,
\begin{equation} \label{dgi}
 \delta_\theta (g_*^{-1}) = g_*^{-1} * \delta_\theta g * g_*^{-1} \,.
\end{equation}
Surprisingly, there is no minus sign on the right hand side. This is 
due to the contribution of the last term in (\ref{var}).

We are now prepared to address the question of variation of the Wilson 
lines under the Seiberg-Witten map. Let us apply the transformation 
$\delta_\theta$ to equation $g_i(x)=g(x)M_i$. The result reads,
$$
 i\lambda(\delta \theta, A) *g_i =
  i \lambda(\delta \theta, A) *gM_i + g \delta_\theta M_i \, .
$$
Thus, we conclude that $\delta_\theta M_i=0$ and monodromies $M_i$ 
are $\theta$-independent, and so are the Wilson lines $W_i(A)$.

Finally, we return to the question of existence of solution $g(x)$ of 
the equation (\ref{Ag}). Clearly, such solutions exist for $\theta=0$.
We can use these solutions as initial conditions in the pair of 
differential equations (\ref{dA}) and (\ref{dg}). Choosing a path 
between $\theta_0=0$ and some other value  $\theta_1$, one can, at 
least formally, obtain solutions of equation (\ref{Ag}) for
$\theta=\theta_1$ by solving equations (\ref{dA}) and (\ref{dg}). 
In general, these solutions depend on the path between $\theta_0$ and 
$\theta_1$. In other words, two infinitesimal transformations 
$\delta^1_\theta$ and $\delta^2_\theta$ may have a nonvanishing 
commutator. This commutator applied to gauge fields $A$ was first
computed in \cite{AK}. We obtain an elegant formula for the variation 
of $g$,
$$ 
[\delta_\theta^1 , \delta_\theta^2 ] \, g = \frac 1{16}
  \delta\theta^{kl}_1 \delta\theta^{mn}_2 \,
 (\partial_l A_n * \partial_m A_k -\partial_m A_k * \partial_l A_n ) \,.
$$
There is no obvious reason for the right hand side to vanish even in 
the abelian case for $\theta \neq 0$.

\section*{Appendix}
Our goal is to show that transformation (\ref{dg}) is consistent
with formula (\ref{Ag}) provided that eq.~(\ref{dA}) holds. 
Applying the partial derivative $\partial_i$ to equation(\ref{dg}) and 
then using twice that $F=0$, we obtain 
\begin{eqnarray*} 
 && \delta_\theta (\partial_i g) = \\
 && = \frac 1{4i} \delta\theta^{kl} ( \partial_i A_k * A_l  
  + A_k * \partial_i A_l + i A_k * A_l * A_i ) * g \\ 
 && = \frac 1{4i} \delta\theta^{kl} ( \partial_k A_i * A_l  
  + A_k * \partial_l A_i + i A_i * A_k * A_l ) * g \,.
\end{eqnarray*}
Taking into account the antisymmetry of $\theta^{ij}$, we
can rewrite this expression as follows:   
\begin{eqnarray*} 
&& \delta_\theta (\partial_i g) = \frac 1{4i} 
 \delta\theta^{kl} ( 2\partial_k A_i * A_l + \partial_l A_i * A_k \\
 && \phantom{=} + A_k * \partial_l A_i + i A_i * A_k * A_l ) * g  \\
 && = -\frac 12 \delta\theta^{kl} \partial_k A_i * \partial_l g +
 i (\delta_\theta A_i) * g + i A_i * \delta_\theta g \\
 && = i \delta_\theta (A_i * g) \,. 
\end{eqnarray*}
Here we used formulas (\ref{dA}) and (\ref{dg}) in the 
third line, and (\ref{Ag}) and (\ref{var}) in the last line.

\vspace{2mm}
{\bf Acknowledgments:} We would like to thank K. Okuyama, 
V. Roubtsov and V. Schomerus for useful discussions. 
A.A. thanks the organizers and the participants of
the conference on Noncommutative Gauge Theory 
(Leiden, November 1999) for inspiring discussions.
The visit of A.B. to the Institute for Theoretical Physics,
Uppsala University was supported by the grant INTAS 96-196 and by the 
grant VISBY-380 of the Swedish Institute.

\def\baselinestretch{1.0}

\widetext

\end{document}